\documentstyle[12pt]{article}
\begin{document}
\begin{titlepage}
\begin{flushright}
IC/91/86
\end{flushright}
\vspace{0.4cm}

\begin{center}
{\small International Atomic Energy Agency\\
\vspace{0.2cm}

and\\
\vspace{0.2cm}

United Nations Educational Scientific and Cultural Organization}\\
\vspace{0.4cm}

INTERNATIONAL CENTRE FOR THEORETICAL PHYSICS


\vspace{1.5cm}

{\bf NONABELIAN $N=2$ SUPERSTRINGS: HAMILTONIAN STRUCTURE}
\vspace{1cm}

{\small A.P. Isaev \footnote{Laboratory of Theoretical Physics,
JINR, Dubna, P.O. Box 79, Head Post Office, Moscow, USSR.}\quad
and \quad E.A. Ivanov $^{1}$

International Centre for Theoretical Physics, Trieste, Italy.}

\thispagestyle{empty}

\bigskip\bigskip
ABSTRACT
\end{center}
\vspace{0.4cm}

{\small We examine the Hamiltonian structure of nonabelian
N=2 superstrings models which are the supergroup manifold extensions
of N=2 Green-Schwarz superstring. We find the Kac-Moody and Virasoro type
superalgebras of the relevant constraints and present elements of
the corresponding quantum theory. A comparison with the type IIA
Green-Schwarz superstring moving in a general curved 10-d supergravity
background is also given. We find that nonabelian superstrings (for d=10)
present a particular case of this general system corresponding to a special
choices of the background.}
\vspace{2cm}

\begin{center}
{MIRAMARE -- TRIESTE} \vspace{0.2cm}

{\small April 1991}

\end{center}
\vfill






%


\end{titlepage}
\newpage

\section{Introduction}
\setcounter{equation}0

In the second paper devoted to nonabelian $N=2$ superstrings
\cite{II}
we present the basic elements of the corresponding
hamiltonian formalism which is a necessary starting point in
constructing the full quantum theory of this system.

In Sect. 2 we recapitulate the known facts about ordinary
Wess-Zumino-Novikov-Witten (WZNW)
sigma models in the hamiltonian approach \cite{W,FT},
following a version of the latter
employed in \cite{AI}.
It has an advantage of admitting a straightforward
extension to the case of nonabelian $N=2$ superstrings we are interested in.
This system is treated in Sect. 3 : we find the relevant hamiltonian
constraints, show how to divide them into the first and second class ones,
establish the structure of the Virasoro and Kac-Moody-type superalgebras
generated by these constraints. In Sect. 4  we compare our model with the
GS superstring moving in a $N=2 \; d=10$ supergravity background and argue that
the former system (for d=10)
presents a special solution for the latter one (corresponding to the type IIA).

  \section{WZNW sigma models as sigma models on $G_{L}\times
  G_{R}/G_{diag}$  coset space. Hamiltonian structure}
  \setcounter{equation}0
In this section we recall the basic facts
about the hamiltonian formulation of
ordinary WZNW sigma models
\cite{AI,W,FT}. Analogous techniques will be applied in
Sect. 3 to study the hamiltonian structure of nonabelian $N=2$
superstrings.

For reasons to become clear later, we follow the
interpretation of WZNW sigma model on the group
$G$ as a sigma model whose target space is the symmetric coset spase
$G_{L}\times G_{R}/G_{diag}$. This means that one originally deals with
two sets of group coordinates (parametrizing $G_{L}$ and $G_{R}$)
related through right gauge transformations of $G_{diag}$.
In a customary approach,
only one set of the $G$ valued coordinates is introduced, which
corresponds to choosing a particular gauge with respect to the right
gauge $G_{diag}$ transformations.

We start by writing the general action for the WZNW sigma models
\cite{W,KZ} coupled to
$2D$ gravity
\begin{equation}
A=l_{I} \{  \int d^{2}\xi \sqrt{-g} g^{ab} Tr(\omega_{a} \omega_{b})-
2/3 \int d^{3}\xi \; \varepsilon^{abc} Tr(\omega_{a} \omega_{b} \omega_{c}) \} ,
\label{v1}
\end{equation}
where $l_{I}$ is a coupling constant,
$\omega_{a}d\xi^{a}=U^{-1}\partial_{a}Ud\xi^{a}-$ is a left-invariant 1-form
and $U=exp(x^{\mu}(\xi^{0},\xi^{1})R_{\mu})$ is a two-dimensional
matrix field taking values in some Lie group with the algebra
\begin{equation}
[R_{\mu},R_{\nu}]=t_{\mu \nu }^{\lambda}R_{\lambda},\;\;\;
Tr(R_{\mu}R_{\nu})=-\eta_{\mu\nu}
\label{v2}
\end{equation}
(for further notations  see our previous paper \cite{II} ).

Let us make manifest the  $G_{L}\times G_{R}/G_{diag}$ coset structure of
the WZNW sigma model with the action (\ref{v1}). To this end we put
\begin{equation}
U=U_{1}U_{2}^{-1}\; , \;\omega_{a}=U_{2}\omega_{a}^{-}U_{2}^{-1} \equiv
U_{2}(\omega_{a}^{1}-\omega_{a}^{2})U_{2}^{-1}\; ,  \;
\omega_{a}^{j}=U_{j}^{-1}\partial_{a}U_{j}, \label{v3}
\end{equation}
where $U_{j}=\exp (x^{j\mu}R_{\mu})$ belongs to $G_{L}$ if $j=1$ and to
$G_{R}$ if $j=2$ (for convenience we have related
these groups to the same set of generators
$R_{\mu})$. Substituting (\ref{v3}) into (\ref{v1}) we see that the action
(defined now on the whole group $G_{L}\otimes G_{R}$) is trivially
invariant under the right gauge $G_{diag}$ transformations
\begin{equation}
U_{1}(\xi)\;\rightarrow \; U_{1}(\xi)V(\xi) \; ,\; U_{2}(\xi)\;\rightarrow \;
U_{2}(\xi)V(\xi) \; , \; V(\xi)\in G_{diag} \label{v4}
\end{equation}
which may be chosen to gauge away half of the original group coordinates. This
explains why we can interpret the above WZNW sigma model as a sigma model
on the coset space $G_{L}\otimes G_{R}/G_{diag}$.

The variation of the action (\ref{v1}) can be represented as
\begin{eqnarray}
\delta A&=&(-2l_{I})\int d^{2}\xi \{-1/2\delta (\sqrt{-g} g^{ab})
Tr(\omega_{a}^{-} \omega_{b}^{-})+ \nonumber \\
&+&Tr[(\partial_{a}(P_{-}^{ab}\omega_{b}^{1}-
P_{+}^{ab}\omega_{b}^{2})-P_{+}^{ab}[\omega_{a}^{1},\omega_{b}^{2}])
(\omega^{1}-\omega^{2} )]\}, \label{v5}
\end{eqnarray}
where $\omega^{j}=U_{j}^{-1}\delta U_{j},  P_{\pm}^{ab}=\sqrt{-g}g^{ab}\pm
\varepsilon^{ab}$. It yields the equations of motion in the form
\begin{equation}
\partial_{a}(P_{-}^{ab}\omega_{b}^{1\mu}-P_{+}^{ab}\omega_{b}^{2\mu})+
t_{\lambda\nu}^{\mu}P_{+}^{ab}\omega_{a}^{1\nu}\omega_{b}^{2\lambda}=0,
\label{v6}
\end{equation}
\begin{equation}
\omega_{a}^{-\mu}\omega_{b\mu}^{-}-1/2g_{ab}g^{cd}\omega_{c}^{-\mu}
\omega_{d\mu}^{-}=0. \label{v7}
\end{equation}

It is worth mentioning that eqs. (\ref{v6}),(\ref{v7}) look very similar
to the equations  of motion of nonabelian
superstrings (see eqs. (3.44),(3.46) in \cite{II}). Eqs. (\ref{v6}) can
be also brought into the form of the current conservation laws
\begin{eqnarray}
\partial_{a}(P_{-}^{ab}U^{-1}\partial_{b}U)&=& 0 \nonumber \\
\partial_{a}(P_{+}^{ab} \partial_{b} UU^{-1})&=&0 \label{v8}
\end{eqnarray}
which may be derived directly from the variation (\ref{v5}) if we
rewrite it as
\begin{eqnarray}
\delta A & = & (-2l_{I})\{\int d^{2}\xi [-1/2\delta(\sqrt{-g} g^{ab})
Tr(\omega_{a}^{-} \omega_{b}^{-}) \nonumber  \\
 & + & Tr[(\partial_{a}(P_{+}^{ab}\partial_{b}
UU^{-1})\delta U_{1}U_{1}^{-1}-\partial_{a}(P_{-}^{ab}U^{-1}\partial_{b}U)
\delta U_{2}U_{2}^{-1})]\}. \label{v9}
\end{eqnarray}
From this representation for $\delta A$ it also follows that $A$ is invariant
($\delta A=0$) if $\delta(\sqrt{-g}g^{ab})=0$ and  $\delta U_{j}$
satisfy the constraints
\begin{equation}
P_{+}^{ab}\partial_{b}(\delta U_{2}U_{2}^{-1})=P_{-}^{ab}\partial_{b}
(\delta U_{1}U_{1}^{-1})=0  \label{v9a}
\end{equation}
or, in the conformal gauge $\sqrt{-g}g^{ab}\sim diag(1,-1) \; , \; \xi_{\pm}=
1/2(\xi^{0}\pm \xi^{1})$ ,
\begin{equation}
\partial_{+}(\delta U_{2}U_{2}^{-1})=\partial_{-}(\delta U_{1}
U_{1}^{-1})=0.    \nonumber
\end{equation}
This amounts to the well-known property that the WZNW model possesses a symmetry
under the two sets of local Kac-Moody transformations \cite{W,KZ,DVR}
\begin{equation}
U_{j}\rightarrow V_{j}(\xi)U_{j} \; , \; \partial_{+}V_{2}=\partial_{-}V_{1}=0.
\label{v10}
\end{equation}

Eqs. (\ref{v8}) express the fact of conservation of the corresponding currents
\begin{equation}
\bar{J}^{1a}=-2l_{I}P_{+}^{ab}\partial_{b}UU^{-1} ,
\bar{J}^{2a}=2l_{I}P_{-}^{ab}U^{-1}\partial_{b}U \label{v10a} .
\end{equation}

It is known that the general solutions of the eqs. (\ref{v8}) have the
form
\begin{equation}
U(\xi^{0},\xi^{1})=U_{1}U_{2 }^{-1}=
\tilde{U}_{1}(\xi_{+}) \tilde{U}_{2}^{-1}(\xi_{-}),
\label{v11}
\end{equation}
where $\tilde{U}_{j}$ are arbitrary
G-valued functions. In terms of the chiral variables
we, correspondingly, have
\begin{equation}
U_{1}(\xi^{0},\xi^{1})=\tilde{U}_{1}(\xi_{+})V(\xi^{0},\xi^{1}) \; , \;
U_{2}(\xi^{0},\xi^{1})=\tilde{U}_{2}(\xi_{-})V(\xi^{0},\xi^{1}),
\label{v12}
\end{equation}
where $V(\xi^{0},\xi^{1})$ is an arbitrary gauge $G_{diag}$ transformation.

An unambiguous method of constructing the hamiltonian formalism for WZNW
models has been worked out by Witten \cite{W}.
This method is based upon the possibility to
introduce a nondegenerate symplectic form with
the help of which one can define the canonical Poisson
brackets. Unfortunately, in the case of nonabelian superstrings \cite{II}
the corresponding symplectic form is degenerate
and so Witten's method in its original form fails to be efficient.
 This is the reason why we prefer here a different hamiltonian
approach used in
\cite{AI}. Moreover, we will treat the
variables $U_{1}$ and $U_{2}$ as independent variables connected by
the gauge transformation (\ref{v4}) (normally, one chooses as the independent
variable the principal chiral field $U=U_{1}U_{2}^{-1}$ ). This will
allow us to keep manifest the $G_{L} \otimes G_{R}/G_{diag}$ coset structure
inherent in the WZNW sigma model and to perform a straightforward extension
to the case of nonabelian superstrings.

First of all , we rewrite the action (\ref{v1})
as a two-dimensional integral by introducing auxiliary field $B_{\mu \nu}$
according to
$$
Tr(\omega^{j} \wedge \omega^{j} \wedge \omega^{j} )=
d(dx^{\mu} dx^{\nu} B_{\mu \nu}),
$$
where it is convenient to choose
\begin{equation}
d(dx^{\mu} dx^{\nu} B_{\mu \nu}) =
d \; Tr(\omega^{j} \wedge \omega^{j} B^{j}).
\label{v13}
\end{equation}
Of course, the fields $B^{j}$ are well defined only
locally \cite{W,VAS} (these
can be explicitly expressed in terms of local coordinates on $G$).
Then, for the action (\ref{v1}) we get the expression
\begin{eqnarray}
\frac{1}{2l_{I}} A & = & - \frac{1}{2l_{I}} \int d^{2} \xi {\cal L}(\xi) =
\int d^{2} \xi \{ -\frac{1}{2}
\sqrt{-g}g^{ab}\omega^{-\mu}_{a} \omega^{-}_{b\mu} +
\varepsilon^{ab} \omega^{1\mu}_{a} \omega^{2}_{b\mu} \nonumber  \\
 & + & \frac{1}{6} t_{\lambda \nu \mu} \varepsilon^{ab}
(\omega^{1\mu}_{a} \omega^{1\nu}_{b} B^{1 \lambda} -\omega^{2\mu}_{a}
\omega^{2\nu}_{b} B^{2\lambda} ) \},
\label{v14}
\end{eqnarray}
where $t_{\lambda \mu \nu }=-\eta_{\lambda \rho }t^{\rho}_{\mu \nu }$.
The action (\ref{v14}) allows one to deal with the ordinary $2D$ Lagrangian ${\cal L}$.
This Lagrangian gives rise to the following expression for the canonical
momentum
\begin{equation}
P^{j}_{\mu}=\frac{\partial {\cal L}}{\partial \partial_{0} x^{j\mu} } =
2l_{I}(-)^{j} \{ P^{0b}_{j} \omega^{-}_{b\rho} -
\varepsilon^{0b} \omega^{j}_{b\rho}-
\frac{1}{3} \varepsilon^{0b} t_{\lambda \nu \rho }
\omega^{j\nu}_{b}B^{j\lambda} \}E^{j\rho}_{\mu}.
\label{v15}
\end{equation}
Here we use the notations $\omega^{j\rho}_{a}=\partial_{a} x^{j\mu}
E^{j\rho}_{\mu}, P^{ab}_{1}=P^{ab}_{+},P^{ab}_{2}=P^{ab}_{-}$. We see
that the momenta (\ref{v15}) are the functions of $B^{j\mu}$
and thus are also well defined only locally. Nevertheless, using (\ref{v15})
one can introduce the globally defined chiral combinations of momenta and
fields $B^{j\mu}$ which, on the one hand, are the linear combination of the
Cartan's forms and, on the other, form the Kac-Moody-type algebras
with respect to the Poisson brackets (see below)
\begin{eqnarray}
J^{j}_{\mu}&=&-\tilde{E}^{j\rho}_{\mu}P^{j}_{\rho}+
2l_{I}(-)^{j} \varepsilon^{0b} (\frac{1}{3} t_{\lambda \nu \mu }
\omega^{j\nu}_{b}B^{j\lambda}-\omega^{j}_{b\mu})=  \nonumber \\
&=&2l_{I}(-)^{j}(P^{0b}_{-}\omega^{1}_{b\mu}-
P^{0b}_{+}\omega^{2}_{b\mu}). \label{v16}
\end{eqnarray}
Here we have introduced the inverse
matrices $\tilde{E}^{j\mu}_{\rho} \; , \tilde{E}^{j\mu}_{\nu}
E^{j\rho}_{\mu}=\delta^{\rho}_{\nu}$.
Then, using eq. (\ref{v16}) we conclude that
\begin{equation}
J^{1}_{\mu}+J^{2}_{\mu}\approx 0 .
\label{v17}
\end{equation}
In the hamiltonian formalism this gives us the constraint
on the hamiltonian variables which relates the $U_{1}$ and
$U_{2}$ degrees of
freedom. This constraint generates the $G_{diag}$ gauge
transformations (\ref{v4}),(\ref{v12}).

First of all we rewrite eq. (\ref{v7}) in terms of the hamiltonian
variables (\ref{v16})
\begin{equation}
2l_{I} {\cal A}^{j}=\frac{1}{4} (J^{j}_{\mu}+4l_{I}(-)^{j}\omega_{1\mu}^{j})
(J^{j\mu}+4l_{I}(-)^{j}\omega_{1}^{j\mu}) \approx 0.
\label{v18}
\end{equation}

The quantities ${\cal A}^{j} (j=1,2)$ are the holomorphic and antiholomorphic
components of the energy-momentum
tensor. Using the definition
of the canonical equal time Poisson brackets
\begin{equation}
\{x^{j\mu}(\xi),P^{i}_{\nu}(\xi ')\} =\delta^{\mu}_{\nu}
\delta^{ij} \delta (\xi -\xi '),
\label{v19}
\end{equation}
after a straightforward calculation we obtain the following algebra of
 the variables $J^{j}_{\mu} ,\omega^{j}_{1\nu}$
\begin{eqnarray}
\{J^{j}_{\mu}(\xi),J^{j}_{\nu}(\xi')\} & = & t^{\lambda}_{\mu \nu}
J^{j}_{\lambda}(\xi) \delta (\xi -\xi ')-(-)^{j}4l_{I}\delta '
(\xi '-\xi ) \eta_{\mu \nu}, \nonumber \\
\{J^{j}_{\mu}(\xi),\omega^{j}_{1\nu} \} & = & \delta (\xi -\xi ')
t^{\lambda}_{\mu \nu} \omega^{j}_{1\lambda} + \delta '(\xi '-\xi )
\eta_{\mu \nu}, \nonumber \\
\{\omega^{j}_{1\nu} , \omega^{j}_{1\mu} \} & = & 0.
\label{v20}
\end{eqnarray}
Thus, the quantities $ \; J^{j}_{\mu }(\xi ) \; $
generate two Kac-Moody algebras and,
quantizing them  ($\{ .,. \}  \rightarrow -i[ .,. ]$), we arrive (choosing
an appropriate representation of $G$) at the
quantization of the parameter $l_{I}$ :
$l_{I}=-\frac{N}{16\pi } , \; \; N \in Z$. To complete an analogy with
the nonabelian superstring hamiltonian structure which will be exposed
in the next Section, we give the expressions for the
temporal components of the Kac-Moody currents
(\ref{v10a}) in terms of the hamiltonian
variables $(J^{j\mu},\omega^{j\mu}_{1})$
\begin{equation}
\bar{J}^{j}(\xi )=U_{j}(J^{j\mu }+4l_{I}(-)^{j}\omega ^{j\mu }_{1})
R_{\mu }U^{-1}_{j}.
\label{v21}
\end{equation}
Indeed, using the definition (\ref{v16}), we find that in terms of the
principal chiral field $U=U_{1}U_{2}^{-1}$ the quantities
$\bar{J}^{j}$ are represented as
\begin{eqnarray}
\bar{J}^{1}(\xi )&=&-2l_{I}P^{0b}_{+}(\partial_{b} U)U^{-1} , \nonumber \\
\bar{J}^{2}(\xi )&=&2l_{I}P^{0b}_{-} U^{-1}(\partial_{b} U) \label{v22}
\end{eqnarray}
which are just the temporal components of the conserved chiral
currents (\ref{v10a}). These components are often denoted
as $\bar{J}^{1}=J , \bar{J}^{2}=\bar{J}$ \cite{W,KZ,GW}.
 The charges corresponding to $\bar{J}^{1},\bar{J}^{2}$ generate the
left and right global symmetry transformations of the principal
 chiral field $U$ (the global limit of transformations (\ref{v10}))
 \begin{equation}
 U \rightarrow G_{L}UG_{R}^{-1} \; \; (U_{1} \rightarrow G_{L}U_{1} ,
 U_{2} \rightarrow G_{R}U_{2} ).
 \label{v23}
 \end{equation}
 In the conformal gauge the conservation laws (\ref{v8})
 and the components (\ref{v22})
 take the form
 \begin{equation}
 \partial_{-}\bar{J}^{1}=\partial_{+}\bar{J}^{2}=0
 \label{v22a}
 \end{equation}
 \begin{equation}
 \bar{J}^{1}=-2l_{I}(\partial_{+}UU^{-1}),\bar{J}^{2}=
 2l_{I}(U^{-1}\partial_{-}U)
 \label{v22b}
 \end{equation}
 Using (\ref{v19}) and (\ref{v20}) we get the following
 commutation relation for $\bar{J}^{j}$ \cite{W,KZ,FT}
 \begin{equation}
 \{\bar{J}^{j}_{\mu}(\xi ) , \bar{J}^{j}_{\nu}(\xi ')\} =
 \-t^{\lambda}_{\mu \nu}\bar{J}^{j}_{\lambda }(\xi ) \delta (\xi -\xi ')+
 (-)^{j}4l_{I}\delta '(\xi '-\xi )\eta_{\mu \nu} .
 \label{v24}
 \end{equation}
 and
 \begin{equation}
 \{\bar{J}^{j}_{\mu}(\xi ) , \bar{J}^{i}_{\nu}(\xi ')\} = 0\; ,\; j\neq i
 \label{v25}
 \end{equation}
 Thus $\bar{J}^{j}(\xi )$ represent two mutually commuting Kac-Moody
 algebras which are just those which generate the transformations
 (\ref{v10}). The energy-momentum components $A^{j}$ (\ref{v18})
 can also be expressed in terms
of $\bar{J}^{j}_{\mu}(\xi )$ (\ref{v21}), after
 that they acquire the familiar Sugawara form
(no summation over j!)
 \begin{equation}
 {\cal A}^{j}=-\frac{1}{4} Tr(\bar{J}^{j} \bar{J}^{j})=
 \frac{1}{4} \bar{J}^{j}_{\mu} \bar{J}^{j \mu}.
 \label{v26}
 \end{equation}
 (recall that in the standard notation \cite{BPZ,KZ} one denotes
 ${\cal A}^{1}=T , {\cal A}^{2}=\bar{T}$).
 The quantum versions of these quantities
 were considered in many papers \cite{W,KZ,GW,GO}.

 In conclusion we stress that the aim of this section was to perform a
 consistent hamiltonian consideration of the standard WZNW sigma model
 from a non-standard point of view, namely as a sigma model
 on the coset spase $G_{L}\otimes G_{R}/ G_{diag}$.
 In the next section we will explore in the same
 manner the hamiltonian structure of the nonabelian superstrings
 described by a sigma model on the coset space
 $G^{*} \otimes G/ G_{+}$  \cite{II} .

 \section{Hamiltonian structure of nonabelian $N=2$ superstring theory.
  Generalized Siegel's algebra}
 \setcounter{equation}0

 In \cite{II} we have found that the nonabelian  $N=2$
 superstrings are described by the generic action
 \begin{eqnarray}
 A&=&l_{I} \{ \int d^{2} \xi (-\sqrt{-g}g^{ab}\omega^{-\mu }_{a}
 \omega ^{-}_{b\mu }+
 \varepsilon ^{ab} \omega ^{-}_{a\mu } \omega ^{+\mu }_{b})- \nonumber \\
 &-&\frac{2}{3} \int d^{3} \xi \varepsilon^{abc} Str(\omega ^{1}_{a}
 \omega ^{1}_{b} \omega ^{1}_{c} -\omega ^{2}_{a} \omega ^{2}_{b}
 \omega ^{2}_{c}) \} \label{u1}
 \end{eqnarray}
 where the notation is explained in the previous section and in \cite{II}.
 We recall that the model with the action (\ref{u1}) is
 the WZNW sigma model defined on the nonsymmetric target superspace
 $G_{1} \otimes G_{2}/G_{+}$ where $G_{1}$ and $G_{2}$ are the supergroups
 generated by two mutually commuting superalgebras dual
to each other in Cartan's sense \cite{KNB}
 $( \mu ,\nu ,\lambda ,...=1,2,...,d; \alpha ,\beta ,\gamma ,...=1,2,...,D)$
 \begin{eqnarray}
 \{ S^{j}_{\alpha },S^{j}_{\beta } \}=(-)^{j}\Gamma^{\mu }_{\alpha \beta}
 R^{j}_{\mu } & , & [ R^{j}_{\mu },S^{j}_{\alpha } ]=C^{\beta }_{\mu \alpha }
 S^{j}_{\beta } ,\nonumber \\
\left[ R^{j}_{\mu },R^{j}_{\nu } \right] & = &
 t^{\lambda }_{\mu \nu } R^{j}_{\lambda } . \label{u2}
 \end{eqnarray}
 with the nondegenerate metrics
 \begin{eqnarray}
 Str(R^{j}_{\mu } R^{j}_{\nu })=-\eta _{\mu \nu } &,& Str(S^{j}_{\alpha }
 S^{j}_{\beta })=(-)^{j+1}X_{\alpha \beta }, \nonumber  \\
 Str(R^{j}_{\mu }S^{j}_{\beta })=0.  \label{u3}
 \end{eqnarray}
 Let us note that the supergroups $G_{1}$ and $G_{2}$ play
 for the nonabelian superstring sigma-model the role similar to
 groups $G_{L}$ and $G_{R}$ for WZNW sigma-model (see the previous section).
 For further convenience, the one-forms $\omega ^{j}_{a} d\xi ^{a}$
 (defined on the algebras of
  $G_{1}$ and $G_{2}$) will be related to the
 same superalgebra, $R^{1}_{\mu }=R^{2}_{\mu }=R_{\mu },
 S^{1}_{\beta }=iS^{2}_{\beta }=S_{\beta }$,
 \begin{equation}
 \omega ^{j}_{a} = U^{-1}_{j}\partial_{a} U_{j}=\omega ^{j\mu }_{a} R_{\mu }+
 \omega ^{j\alpha }_{a}S_{\alpha },  \label{u4}
 \end{equation}
 where $U_{j}= \exp (\frac{i}{2} x^{j\mu } R_{\mu }-(-i)^{j}\theta ^{j\alpha}
 S_{\alpha })$
 \begin{eqnarray}
 \{ S_{\alpha },S_{\beta } \}=-\Gamma ^{\mu }_{\alpha \beta} R_{\mu }&,&
 \left[ R_{\mu },S_{\alpha }\right]
 =C^{\beta }_{\mu \alpha}S_{\beta }, \nonumber \\
 \left[ R_{\mu },R_{\nu }\right]
 &=& t^{\lambda }_{\mu \nu}R_{\lambda }. \label{u5}
 \end{eqnarray}
 Let us mention here that the definition
 of the coefficients $\omega ^{jA}_{b}$
 of the Maurer-Cartan forms in this paper is slightly different from
 the one adopted in \cite{II}. Namely, all the
 coefficients are the same except for
 $\omega ^{2\alpha }_{a}$ (in \cite{II} we have used
 $-i\omega ^{2\alpha }_{a}$ ). We make this redefinition
  to avoid undesirable factors $i$ in the subsequent formulas.

 It is easy to see that if we put all the fermionic forms in the action
 (\ref{u1}) equal to zero
 ($\omega^{j\alpha }_{a}=0$) , we arrive just at the
 action (\ref{v1}). Thus the model based upon the action (\ref{u1}) is a
 superextension of the previously considered WZNW sigma model coupled
 to two dimensional gravity.

 Further in this section we will often use the supernotation
 which is much more concise. Instead of $R_{\mu},S_{\alpha }$ we
 introduce the generic notation $T_{A}$, where $A$ runs over
 all possible indices $\mu $ and $\alpha $. We introduce also
 the supermetric $X_{AB}=Str(T_{A}T_{B})=(-\eta_{\mu \nu },
 X_{\alpha \beta})$. Then the commutation relations (\ref{u5})
 can be rewritten in the condensed form ($t^{\alpha }_{\mu \beta}=
 C^{\alpha }_{\mu \beta }, t^{\mu}_{\alpha \beta }=
 -\Gamma ^{\mu}_{\alpha \beta }$)
 \begin{equation}
 [T_{A},T_{B}]=t^{C}_{AB}T_{C}    \label{u6}
 \end{equation}
 where the structure constants $t^{A}_{BC}$ obey the Jacobi
 identity
 \begin{equation}
 (-)^{BD}t^{C}_{AB}t^{E}_{DC}+(-)^{AB}t^{C}_{DA}t^{E}_{BC}+
 (-)^{DA}t^{C}_{BD}t^{E}_{AC}=0  \label{u6a}
 \end{equation}
 and $t^{A}_{BC}=0$ if $(A)+(B)+(C)\neq 0$. Here $(A)$ is the Grassman
 parity of the generator $T_{A}$: $(A)=0\; (mod \; 2)$ if
 $A=\mu $ and $(A)=1\; (mod \; 2)$ if $A=\alpha $. It is also useful to
 introduce the structure constants with the lowered indices
 $t_{ABC}=X_{AD}t^{D}_{BC}$. These constants possess the following
 symmetry properties
 \begin{equation}
 t_{ABC}=(-)^{BC+1}t_{ACB}=(-)^{A}t_{BCA}  \label{u7}
 \end{equation}
 Here and in eqs. (\ref{u6a}) we put $(-)^{A}=(-1)^{(A)}$ , etc. . Using
 the supernotation one can rewrite the one-forms $\omega^{j}=
 U^{-1}_{j}dU_{j}=\omega^{j}_{a} d\xi ^{a}$ (see equation (\ref{u4}) )
 as follows
 \begin{equation}
 \omega^{j}=\omega^{jA }_{a} T_{A} d\xi ^{a}=(\partial _{a}z^{jM}
 E^{jA}_{M})T_{A}d\xi ^{a}  \label{u8}
 \end{equation}
 where $z^{jM}=(x^{j\mu }, \theta ^{j\alpha })$ are the even and odd parameters
 of the supergroups $G_{j}$ and $E^{jA}_{M}$ are the one-forms
 coefficients which can be identified with supervelbeins. The
 Maurer-Cartan equations (see (3.15) in \cite{II} ) can be rewritten as
 \begin{equation}
 \partial_{M} E^{jA}_{N}-(-)^{MN}\partial_{N} E^{jA}_{M}=
 (-)^{CN} E^{jC}_{M}E^{jB}_{N}t^{A}_{BC} \label{u9}
 \end{equation}

 Let us return to studying the system with the action (\ref{u1}).
 Like in the case considered in the previous section it is not
 so trivial to explore the hamiltonian structure for this action because of
 nonlocality of the WZ term in eq. (\ref{u1}).
 One might try to treat such an action using Witten's approach
 \cite{W} . However, as was mentioned in previous Section,
 this approach  does not apply
 straightforwardly to our model, because the corresponding
 symplectic form is degenerate and therefore (as we will see)
 there appear constraints
 on the hamiltonian variables. By this reason, it proves more fruitful
 to use the techniques displayed in the previous section. To this end,
 it is necessary to rewrite WZ term as a two-dimensional
 integral by introducing an auxiliary fields $B^{j}$ (see eq. (\ref{v13}))
 defined by the equation (which is valid only locally)
 \begin{equation}
 Str(\omega^{j}\wedge \omega^{j}\wedge \omega^{j})=
 d\; Str(\omega^{j}\wedge \omega^{j}\wedge B^{j}).  \label{u10}
 \end{equation}

 Now the action (\ref{u1}) is rewritten in the form
 \begin{eqnarray}
 A & = & - \int d^{2} \xi {\cal L} (\xi )= l_{I}
 \int d^{2} \xi \{ -\sqrt{-g} g^{ab}
 \omega^{-\mu }_{a} \omega^{-}_{b\mu }   \nonumber   \\
 &+& 2\varepsilon^{ab}
 \omega^{1 \mu }_{a}\omega^{2}_{b\mu }+\frac{1}{3} \varepsilon^{ab}
 t_{ABC}(\omega^{1C}_{a}\omega^{1B}_{b} B^{1A}-
 \omega^{2C}_{a}\omega^{2B}_{b} B^{2A}) \}. \label{u11}
 \end{eqnarray}
 Here $B^{j}=B^{jA}T_{A}$ and $\cal{L}(\xi )$ is the Lagrangian of our
 model. This Lagrangian results in the following expression for
 the canonical momentum
 \begin{eqnarray}
 P^{j}_{M} &=& \frac{\partial {\cal L}}{\partial \partial_{0} {\rm z}^{jM}}=
 (-2l_{I}) \{ (-)^{j}P^{0b}_{j} \omega^{-}_{b\mu }E^{j\mu }_{M}-
 (-)^{j}\varepsilon^{0b}\omega^{j}_{b\mu }E^{j\mu }_{M}   \nonumber \\
 &-& (-)^{j}\frac{1}{3} \varepsilon^{0b} t_{ABC} E^{jC}_{M}
 \omega^{jB}_{b} B^{jA} \},   \label{u12}
 \end{eqnarray}
 where $P^{ab}_{j}$ are the projection operators (for notations see eqs.
 (\ref{v15})). The canonical momentum $P^{j}_{M}$ and the one-forms
 $\omega^{j}_{1}$ form the complete set of the hamiltonian variables
 for our model. Unfortunately, the Lagrangian ${\cal L}$ is singular and
 therefore velocities $\dot{z}^{jK}$ (or the left-invariant forms
 $\omega^{jK}_{0}$) are not expressed in terms of the hamiltonian
 variables $P^{j}_{M}$ and $\omega^{j}_{1}$.
 That is why we will have constraints on these variables. We will disscuss
 these constraints below.

 Further for convenience we will reserve the letters $K,L,M,N,...$ for
 the indices of the group parameters $z^{jK}$ and for the low indices
 of the supervielbeins $E^{jA}_{K}$ while the letters
 $A,B,C,D,...$ for the indices of the group generators $T_{A}$
 and for the upper indices of $E^{jA}_{K}$ . In other words $K,L,M,N,...$
 are world indices while $A,B,C,...$ are the tangent space indeces.
 Keeping in mind this convention we denote the inverse supermatrices for
 $E^{jA}_{ K}$ as $E^{jM}_{ A}$ i.e.
 \begin{equation}
 E^{jM}_{ A}E^{jB}_{ M}=\delta^{B}_{A} \; , \;
 E^{jA}_{ M}E^{jK}_{ A}=\delta^{K}_{M}
 (no \; summation \; over \; j) .   \label{u13}
 \end{equation}

 Now let us note that the dependence of $P^{j}_{M}$ (\ref{u12}) on
 the fields $B^{jA}$ means that the expression (\ref{u12}) is
 ill-defined globally. Using the inverse  supermatrices $E^{jM}_{ A}$
 and eq. (\ref{u12}) we can introduce the quantities
 which are linear combinations of the coefficients of the left-invariant
 forms (\ref{u8}) and by construction are independent of $B^{jA}$ (cf.
 eq. (\ref{v16}))
 \begin{eqnarray}
 J^{j}_{C} &=& -E^{jM}_{ C}P^{j}_{M} +2l_{I} (-)^{j} (1/3t_{ABC}
 \omega^{jB}_{1} B^{jA}+\omega^{jB}_{1} X_{BC})   \nonumber  \\
 &=& 2l_{I} (-)^{j}
 [ (P^{0b}_{-}\omega^{1\mu }_{b}-P^{0b}_{+}\omega^{2\mu }_{b})\eta_{\mu C} +
 \omega^{j\alpha }_{1} X_{\alpha C} ] \label{u14}
 \end{eqnarray}

 These quantities are related to the conserved currents appearing in
 the nonabelian superstring model (see (3.47) in \cite{II}).
 The independence $J^{j}_{C}$ of $B^{j}$ means
 that the quantities $J^{j}_{C}$ can be defined globally.
 Now, instead of the complete
 set of the hamiltonian variables $ \{ P^{j}_{\mu} , P^{j}_{\alpha} ,
 \omega^{j\mu}_{1} ,\omega^{j\alpha}_{1}
 \; \; (j=1,2) \} $ we introduce a new complete set
 $\{ J^{j}_{\mu},J^{j}_{\alpha},\omega^{j\mu}_{1} ,\omega^{j\alpha}_{1}
 \; \; (j=1,2) \} $ where $J^{j}_{A}$ are expressed
 as linear combinations of $P^{j}_{M}$ and are well
 defined globally. As we will see below, our choice of the
 quantities (\ref{u14})(instead of $P^{j}_{M}$)
 is also convenient in that they form
 Kac-Moody superalgebras with respect to the Poisson superbrackets.

 From eqs. (\ref{u12}) and (\ref{u14}) one obtains the following
 set of the primary constraints
 \begin{eqnarray}
 {\cal D}^{j}_{\alpha } &=& -E^{jM}_{\alpha }P^{j}_{M}+(-)^{j}\frac{2l_{I}}{3}
 t_{AB\alpha } \omega^{jB}_{1} B^{jA}   \nonumber   \\
 &=& J^{j}_{\alpha }-(-)^{j}2l_{I}\omega^{j\beta }_{1} X_{\beta \alpha }
 \approx 0,  \label{u15}
 \end{eqnarray}
 \begin{equation}
 {\cal J}_{\mu }=J_{\mu }^{1}+J^{2}_{\mu } \approx 0.  \label{u15b}
 \end{equation}
 We have also to add the constraint following from
 the action (\ref{u11}) by varying it with respect to the two-dimensional
 metric $g^{ab}$ (see eq. (\ref{v7}) )
 \begin{equation}
 \omega^{-\mu }_{a}\omega^{-}_{b\mu }-1/2g_{ab}g^{cd}
 \omega^{-\mu }_{c}\omega^{-}_{d\mu }=0  \nonumber
 \end{equation}
 or, in the equivalent form
 \begin{equation}
 {\cal A}^{j} \sim P^{0a}_{j}
 \omega^{-\mu }_{a} P^{0b}_{j} \omega^{-}_{b\mu }=0     \nonumber
 \end{equation}
 In terms of the hamiltonian variables (\ref{u12}), (\ref{u14}) these
 constraints can be rewritten as
 \begin{eqnarray}
 2l_{I}{\cal A}^{j} &=& \frac{1}{4} ((-)^{j}J^{j}_{\mu }
 +4l_{I}\omega_{1\mu }^{j})
 ((-)^{j}J^{j\mu }+4l_{I}\omega^{j\mu }_{1})  \nonumber   \\
 &=& \frac{1}{4}J^{j}_{\mu }J^{j\mu }+2l_{I}(-)^{j}
 (-\partial_{1}z^{jM} P^{j}_{M}-\omega^{j\alpha }_{1}{\cal D}^{j}_{\alpha })
 \approx 0  \label{u15c}
 \end{eqnarray}
 We stress that eq. (\ref{u15c}) is valid only on shell and ${\cal A}^{j}$
 are none other than the components of the energy-momentum tensor
 (see (\ref{v18})). The constraint (\ref{u15b}) (cf. eq. (\ref{v17}))
 is a consequence of our coset construction specific for the action
 (\ref{u1}) : it generates gauge transformations forming
 to the right gauge group $G_{+}$.

 Now, in order to divide the constraints (\ref{u15}) into
 the first and second class ones it is necessary
 to calculate the Poisson brackets of the quantities (\ref{u15}).
 For the variables $P^{j}_{M}(\xi )$ and $z^{jN}(\xi ')$ one
 postulates the following equal time
 canonical Poisson superbrackets
 \begin{equation}
 \{ P^{j}_{M}(\xi ),z^{jN}(\xi ')\}=
 (-)^{N+1}\{ z^{jN}(\xi '),P^{j}_{M}(\xi ) \}=\delta^{N}_{M}\delta (\xi-\xi ')
 \label{u16}
 \end{equation}
 Using the definition (\ref{u10}) of the auxiliary
 fields $B^{j}$ we find that the quantities (\ref{u14})
 form two ($j=1,2$) mutually commuting Kac-Moody superalgebras with
 respect to the Poisson superbrackets (\ref{u16})
 \begin{equation}
 \{J^{j}_{A}(\xi ),J^{j}_{B}(\xi ')\} =t^{C}_{AB}J^{j}_{C}(\xi )
 \delta (\xi -\xi ')+(-)^{j}4l_{I}\delta '(\xi '-\xi )X_{AB}
 \label{u17a}
 \end{equation}
 \begin{equation}
 \{ J^{j}_{A}(\xi ),\omega^{jC}_{1}(\xi ')X_{CB}\}=
 \delta(\xi -\xi ')t^{C}_{AB}\omega^{jD}_{1}(\xi ')X_{DC}+
 \delta '(\xi '-\xi )X_{AB}
 \label{u17b}
 \end{equation}
 Note that $J^{1}_{A}$ and $J^{2}_{A}$ do not generate independent
 superalgebras in view of the constraints (\ref{u15b})
 (treated in the weak sense). We stress that the algebra
 $\{ J_{\mu }=1/2(J^{1}_{\mu }-J^{2}_{\mu }), J^{i}_{\alpha},
 \omega^{1\mu}-\omega^{2\mu} , \omega^{i\alpha} \}$ with respect to the
 Poisson brackets represents a special case
 of the Bergshoeff-Sezgin algebra \cite{BS} .
 Now we can use
 the relations (\ref{u17a}), (\ref{u17b})
 to obtain the following superalgebra of the primary constraints
  (\ref{u15}),(\ref{u15b}),(\ref{u15c})
 \begin{eqnarray}
 \{ {\cal A}^{j}(\xi) ,{\cal A}^{j}(\xi ')\} &=& (-)^{j}({\cal A}^{j}(\xi ')
 \delta '(\xi '-\xi)-
 {\cal A}^{j}(\xi) \delta '(\xi -\xi '))   \label{u18a}  \\
 \{ {\cal A}^{j}(\xi ),{\cal J}_{\mu }(\xi ') \} &=& 0   \label{u18b}  \\
 \{ {\cal J}_{\mu }(\xi ),{\cal J}_{\nu }(\xi ') \} &=&
 t^{\lambda }_{\mu \nu }
 {\cal J}_{\lambda }(\xi ) \delta (\xi -\xi ')  \label{u18c} \\
 \{ {\cal J}_{\mu }(\xi ),{\cal D}^{j}_{\alpha }(\xi ') \} &=&
 C^{\beta }_{\mu \alpha }
 {\cal D}^{j}_{\beta }(\xi )\delta (\xi -\xi ')  \label{u18d}  \\
 \{ {\cal D}^{j}_{\alpha }(\xi ) ,{\cal D}^{j}_{\beta }(\xi ') \} &=&
 (-)^{j+1}
 \Gamma^{\mu }_{\alpha \beta }((-)^{j}J_{\mu }^{j}+4l_{I}\omega^{j}_{1\mu })
 \delta (\xi -\xi ')  \label{u18e}  \\
 \{ {\cal A}^{j}(\xi ), {\cal D}^{j}_{\alpha } (\xi ') \} &=& \frac{1}{4l_{I}}
 ((-)^{j}J^{j\mu }+4l_{I}\omega ^{j\mu }_{1}) C^{\beta }_{\mu \alpha }
 ((-)^{j}{\cal D}^{j}_{\beta } -4l_{I}\omega ^{j}_{1 \beta })
 \delta (\xi -\xi ')  \label{u18f}
 \end{eqnarray}
It is seen from eqs. (\ref{u18a}) - (\ref{u18f}) that the even quantities
${\cal A}^{j} (\xi),{\cal J}_{\mu} (\xi)$ form the closed subalgebra and
thus (\ref{u15b}), (\ref{u15c}) can be naively considered as
the first class constraints
while odd constraintes ${\cal D}^{j}_{\alpha }(\xi ) \approx 0$
at first sight (see
eqs. (\ref{u18d}),(\ref{u18e}),(\ref{u18f})) are the second class.
A more careful analysis shows that if there exist the matrices
$\tilde{\Gamma }$ such that

\begin{equation}
\Gamma^{\mu}_{\alpha\beta} (\tilde{\Gamma}^{\nu})^{\beta\gamma} +
\Gamma^{\nu}_{\alpha\beta} (\tilde{\Gamma}^{\mu})^{\beta\gamma}=
\eta^{\mu\nu} \delta^{\gamma}_{\alpha} ,   \label{u19}
\end{equation}
then the quantities ${\cal D}^{j}_{\alpha}(\xi)$ are definite
combinations of the
second and first class constraints in a full analogy with the ordinary
superstring theory \cite{GS,WS,KRNP}. We recall
here that
the action (\ref{u1}) possesses $\kappa $-supersymmetry
if the matrices $\tilde{\Gamma}^{\mu} $ exist \cite{II}.

To find correct expressions for the first class constraints
(which can be linear combinations of the quantities
${\cal A}^{j},{\cal J}_{\mu }$ and ${\cal D}^{j}_{\alpha}$ ) it is
necessary to examine how they evolve with time. First
of all, let us note that the canonical hamiltonian for our
model (\ref{u1}) can be chosen as
a linear combination of the constraints (\ref{u15})

\begin{equation}
{\cal H} (\xi_{0}) =\int^{2\pi}_{0} d\xi \{
{\cal A}^{j}f_{A}^{j}(\xi_{0} ,\xi )+
{\cal D}^{j}_{\beta}f^{j\beta }_{D} (\xi_{0} ,\xi )+{\cal J}_{\mu }f^{\mu}_{J}
(\xi_{0} ,\xi ) \} .  \label{u20}
\end{equation}
Here the integration goes over the spatial coordinate $\xi_{1}=
\xi $ (for simplicity, we assume that
all functions are periodical, $f(\xi )=f(\xi + 2\pi )$), and
$f^{j}_{A},f^{j\alpha }_{D},f^{\mu }_{J}$ are Lagrange multipliers,
which can be fixed by the condition of compatibility with (\ref{u15}),
(\ref{u15b}) and (\ref{u15c})
\begin{eqnarray}
\dot{{\cal D}}^{j}_{\alpha } &=& \{ {\cal D}^{j}_{\alpha },
{\cal H} \} \approx 0, \label{u21a} \\
\dot{{\cal A}}^{j} &=& \{ {\cal A}^{j}, {\cal H} \} \approx 0,  \label{u21b}  \\
\dot{{\cal J}}_{\mu } &=& \{ {\cal J}_{\mu },
{\cal H} \} \approx 0. \label{u21c}
\end{eqnarray}
Using the commutation relations (\ref{u18a} -\ref{u18f}) one
can show that  eqs. (\ref{u21a} - \ref{u21c}) amount
to the following consistency conditions
\begin{eqnarray}
((-)^{j}J^{j}_{\mu}+4l_{I}\omega^{j}_{1\mu})\Gamma^{\mu}_{\alpha \beta}
(f^{j\beta}_{D}+2l_{I}(-)^{j}f^{j}_{A}\omega^{j\beta}_{1}) \approx &0&
\nonumber \\
((-)^{j}J^{j}_{\mu}+4l_{I}\omega^{j}_{1\mu})\Gamma^{\mu}_{\alpha \beta}
\omega^{j\alpha }_{1}f^{j\beta}_{D} \approx &0&   \label{u22}
\end{eqnarray}
If $\Gamma^{\mu}_{\alpha \beta}$ satisfy (\ref{u19}), then the general
solution to these equations is
\begin{equation}
f^{j\beta}_{D}=-2l_{I}(-)^{j}f^{j}_{A}\omega^{j\beta }_{1}+
(\tilde{\Gamma}^{\nu})^{\beta \gamma}
((-)^{j}J^{j}_{\nu}+4l_{I}\omega^{j}_{1\nu})\tilde{f}^{j}_{\gamma}
\label{u23}
\end{equation}
 Here $\tilde{f}^{j}_{\gamma}$ are arbitrary odd functions. Substituting
 the new expression for the Lagrange multiplier (\ref{u23}) into (\ref{u20})
 we obtain the hamiltonian in the form
 \begin{equation}
 {\cal H}=\int d\xi \{ \tilde{{\cal A}}^{j}f^{j}_{A}+
 {\cal B}^{j\beta }\tilde{f}^{j}_{\beta }
 +{\cal J}_{\mu }f^{\mu }_{J} \}  , \label{u24}
 \end{equation}
 where
 \begin{eqnarray}
 \tilde{{\cal A}}^{j} &=& {\cal A}^{j}-(-)^{j}{\cal D}^{j}_{\beta }
 \omega^{j\beta}_{1}  , \label{u25a} \\
 {\cal B}^{j\beta} &=& (\tilde{\Gamma}^{\mu})^{\alpha \beta}
 ((-)^{j}J^{j}_{\mu}+
 4l_{I}\omega^{j}_{1\mu}){\cal D}^{j}_{\alpha} .  \label{u25b}
 \end{eqnarray}
 It is easy to check that the quantities ${\cal B}^{j\beta}$
 generate, modulo the
 constraints (\ref{u15}), the local fermionic $\kappa$-supersymmetry
 transformations $(\tilde{\omega}^{j}=U^{-1}_{j}\delta U_{j})$
 \begin{eqnarray}
 \tilde{\omega}^{1\mu } &-& \tilde{\omega}^{2\mu }=0 ,  \nonumber  \\
 \tilde{\omega}^{1\alpha } &=& P^{ab}_{+}\tilde{\omega}_{b\mu }^{-}
 (\tilde{\Gamma}^{\mu })^{\alpha \beta }\kappa^{1}_{a\beta}(\xi ), \nonumber \\
 \tilde{\omega}^{2\alpha } &=& P^{ab}_{-}\tilde{\omega}_{b\mu }^{-}
 (\tilde{\Gamma}^{\mu })^{\alpha \beta }\kappa^{2}_{a\beta}
 (\xi ), \label{u26} \\
 \delta(\sqrt{-g}g^{ab}) = &-& 2(P^{da}_{+}P^{cb}_{+} \omega^{1\beta }_{d}
 \delta^{\gamma}_{\beta}\kappa^{1}_{c\gamma}-
 P^{da}_{-}P^{cb}_{-}\omega^{2\beta }_{d}\delta^{\gamma}_{\beta}
 \kappa^{2}_{c\gamma}). \nonumber
 \end{eqnarray}
 New quantities $\tilde{\cal A}^{j},{\cal B}^{j\alpha }$
 and $J_{\mu }$ are the first
 class constraints but this set of constraints is not closed with
 respect to the Poisson brackets and thus must be completed.
 Commuting these generators among themselves leads to an extended algebra
 of the first class constraints which is
 a generalization of Siegel's ${\cal A,B,C,D}$
 superalgebra \cite{WS} appearing in the ordinary superstring theory.

 Now we introduce, besides the forms (\ref{u14}) ,
 the following currents (cf. the currents (\ref{v21}), (\ref{v22}))
 \begin{equation}
 \bar{J}^{j}(\xi)=\bar{J}^{j}_{A}(\xi)X^{AB}T_{B}=
 U_{j}(J^{j}_{A}X^{AB}T_{B}-4l_{I}(-)^{j}U^{-1}_{j}\partial_{1}U_{j})
 U^{-1}_{j}   \label{u27}
 \end{equation}
 Using the field equations following from the action (\ref{u1})
 (see eqs. (3.47) in \cite{II}) we find that $\bar{J}^{j}$
 in the conformal gauge satisfy the equations (cf. (\ref{v8}),
 (\ref{v22a}))
 \begin{eqnarray}
 \partial_{-}(\bar{J}^{1}) &=& 2l_{I}\partial_{1} (U_{1}\omega^{1\beta}_{-}
 S_{\beta}U^{-1}_{1}) \nonumber \\
 \partial_{+}(\bar{J}^{2}) &=& 2l_{I}\partial_{1} (U_{2}\omega^{2\beta}_{+}
 S_{\beta}U^{-1}_{2}) \nonumber
 \end{eqnarray}
 Thus we see that $\bar{J}^{j}$ are the temporal components of the conserved
 currents assotiated with the left and right global supersymmetries. The
 components (\ref{u27}) also form Kac-Moody superalgebras with
 respect to (\ref{u16})
 \begin{eqnarray}
 \{ \bar{J}^{j}_{A}(\xi ),\bar{J}^{j}_{B}(\xi ') \} &=& -t^{C}_{AB}
 \bar{J}^{j}_{C}(\xi) \delta (\xi '-\xi )-(-)^{j}4l_{I}\delta '(\xi '-\xi )
 X_{AB}  \label{u28a}  \\
 \{ \bar{J}^{j}_{A}(\xi) ,\bar{\omega}^{jC }(\xi ')X_{CB} \} &=&
 -t^{C}_{AB}\bar{\omega}^{jD }X_{DC}\delta (\xi '-\xi )+X_{AB}
 \delta '(\xi '-\xi ). \label{u28b}
 \end{eqnarray}
 Here $\bar{\omega}^{jA}T_{A} =\partial_{1}U_{j}U^{-1}_{j}$. Moreover,
 we have
 \begin{eqnarray}
 \{ J^{i}_{A}(\xi) , \bar{J}^{j}_{B}(\xi ') \}=0 &,&
 \{ \bar{J}^{1}_{A}(\xi) , \bar{J}^{2}_{B}(\xi ') \}=0  \nonumber  \\
 \{ J^{1}_{A}(\xi) &,& J^{2}_{B}(\xi ')  \}=0.   \label{u28c}
 \end{eqnarray}
 We see from (\ref{u17a}), (\ref{u28a}) and (\ref{u28c}) that the
 superalgebras $J^{j}_{A}$ and $\bar{J}^{j}_{A} \; (j=1,2)$ constitute a
 direct sum of four mutually commuting Kac-Moody superalgebras.

 Now let us discuss the constraints (\ref{u15c}) , (\ref{u25a}) in more
 detail. Using formulas (\ref{u27}) we can rewrite the energy-momentum
 tensor components ${\cal A}^{j}(\xi )$ as
 \begin{equation}
 {\cal A}^{j}=-\frac{1}{8l_{I}}(\bar{J}^{j}_{A}X^{AB}
 \bar{J}^{j}_{B}-J^{j}_{\alpha}
 X^{\alpha \beta }J^{j}_{\beta} )+ (-)^{j}
 {\cal D}^{j}_{\alpha} \omega^{j\alpha}_{1} ,  \label{u29}
 \end{equation}
 or, for the improved first class constraint
 $\tilde{{\cal A}}^{j}$ (\ref{u25a}), (cf. eqs. (\ref{v26}))
 \begin{eqnarray}
 \tilde{{\cal A}}^{j} &=& -\frac{1}{8l_{I}}
 (\bar{J}^{j}_{A}X^{AB}\bar{J}^{j}_{B}
 -J^{j}_{\alpha}X^{\alpha \beta }J^{j}_{\beta} ) \nonumber \\
 &=& g_{1}\bar{J}^{j}_{A}X^{AB}\bar{J}^{j}_{B}-
 (g_{2}\bar{J}^{j}_{A}X^{AB}\bar{J}^{j}_{B} - g_{3}J^{j}_{\mu}
 X^{\mu \nu }J^{j}_{\nu } ) ,  \label{u30}
 \end{eqnarray}
 where $g_{1}=g_{2}=g_{3}=-1/(8l_{I})$.
 The expression (\ref{u30}) gives us the coset version \cite{GKO}
 of the Sugawara
 representation of the classical
 Virasoro algebra. It is interesting
 to note that eqs. (\ref{u14}), (\ref{u15c}) and (\ref{u29})
 imply
 \begin{equation}
 1/4(J^{j}_{A}X^{AB}J^{j}_{B})-1/4(\bar{J}^{j}_{A}X^{AB}\bar{J}^{j}_{B})=
 -2l_{I}(-)^{j}\partial_{1}z^{jM}P^{j}_{M} ,  \nonumber
 \end{equation}
 where in the right hand side we find the generators of the superstring
 reparametrization. Now one may canonically
 quantize the Kac-Moody superalgebras
 (\ref{u17a}) and (\ref{u28a}) using the substitution $[.,.]_{\pm}=
 i\{ .,. \}$ and following the consideration in refs. \cite{GOW} .
 As a result, we obtain (choosing the appropriate representation
 $V$ of the superalgebra (\ref{u5}))
 that the parameter $l_{I}$ is quantized,
 $l_{I}=-\frac{N}{16\pi }, \; N$ being a positive integer, and the equal time
 commutators for the Fourier components of $J$ and $\bar{J}$ become
 \begin{eqnarray}
 \left[ J^{jn}_{A} , J^{jm}_{B} \right]_{\pm} &=& t^{C}_{AB}
 J^{jn+m}_{C}-(-)^{j}\frac{Nn}{2}
 X_{AB}\delta_{n+m,0} ,  \label{u31a}  \\
 \left[ \bar{J}^{jn}_{A} ,\bar{J}^{jm}_{B} \right]_{\pm} &=&
 t^{C}_{AB}\bar{J}^{jn+m}_{C}-(-)^{j}\frac{Nn}{2}
 X_{AB}\delta_{n+m,0}  ,    \label{u31b}  \\
 \left[ J^{jn}_{A} , \bar{J}^{jm}_{B} \right]_{\pm} &=& 0  ,   \label{u31c}
 \end{eqnarray}
 where $J^{j}_{A}(\xi )=\frac{i}{2\pi}\sum_{n} J^{jn}_{A}\exp (-in\xi ) , \; \;
 \bar{J}^{j}_{A}(\xi )=-\frac{i}{2\pi}\sum_{n} \bar{J}^{jn}_{A}\exp (-in\xi )$.
 Now we can define the vacuum $ \mid 0 > $ such that
 $J^{jn}_{A} \mid 0 > = \bar{J}^{jn}_{A} \mid 0 > =  0 $ for $( n > 0,
 j=1)$ and for $(n < 0, j=2)$. To quantize the first class constraints
 (\ref{u30}) it is necessary, in an entire analogy with the consideration in
 papers \cite{KZ,GKO,GOW}, to pass to the normal ordering in expression (\ref{u30})
 and to properly renormalize the constants $g_{1},g_{2}$ and
 $g_{3}$. Namely, we have to put
 \begin{equation}
 g_{1}=g_{2}=\frac{2\pi}{N+C_{V}}, g_{3}=\frac{2\pi}{N+C_{V^{0}}} \label{u32}
 \end{equation}
 where $C_{V}$ and $C_{V^{0}}$ are the quadratic Casimir
 invariants for the superalgebra (\ref{u5})
 and its even subalgebra (\ref{v2}), respectively,
 \begin{eqnarray}
 (-)^{B}t^{B}_{DC}t^{C}_{AB}=C_{V}X_{DA} &,&
 t^{\mu}_{\nu \lambda }t^{\lambda }_{\rho \mu}
 =C_{V^{0}}X_{\nu \rho} \nonumber \\
 X_{AB} &=& Str(T_{A}T_{B}).   \nonumber
 \end{eqnarray}
 We recall here (see eq. (\ref{u30})) that the classical
 values for $g_{1}, g_{2}$ and $g_{3}$ are $g_{1}=g_{2}=g_{3}=-1/(8l_{I})=
 2\pi /N$ (cf. with (\ref{u32})).
 Only with the choice (\ref{u32}) the
 normally ordered quantities $\tilde{{\cal A}}^{j}$ generate
 the Virasoro algebras
 \begin{eqnarray}
 \left[ \tilde{{\cal A}}^{j}(\xi) , \tilde{{\cal A}}^{j}(\xi ') \right]
 = & i & (-)^{j}\{
 (\tilde{{\cal A}}^{j}(\xi ')\delta '(\xi '-\xi)-\tilde{{\cal A}}^{j}(\xi)
 \delta '(\xi -\xi ') ) \nonumber \\
 - & c & \frac{\pi}{24}(\delta '''(\xi -\xi ')+\delta '(\xi -\xi '))\}
 \label{u33}
 \end{eqnarray}
 where the central charge $c$ is defined according to
 the coset construction approach \cite{GKO} as
 \begin{equation}
 c=\frac{N}{N+C_{V}}Sdim(g)-(\frac{N}{N+C_{V}}Sdim(g)-
 \frac{N}{N+C_{V^{0}}}dim(g^{0}))=\frac{N}{N+C_{V^{0}}}dim(g^{0}) \nonumber
 \end{equation}
 Here $Sdim(g)=dim(g^{0})-dim(g^{1}) = d-D$ ($g^{0}$ and $g^{1}$ are
 the even and odd sectors of the superalgebra (\ref{u5}),
 respectively).

 To close this section, we would like to stress that the
 knowledge of the central charge is important for
 drawing information about the vanishing of the conformal anomalies
 and about critical dimensions of our nonabelian $N=2$ superstring
 models. However, the existence of the first and second class constraints
 relating the hamiltonian variables requires a more careful treatment of the
 quantum theory (e.g. , it is necesary to introduce harmonics and
 ghosts \cite{KRNP}) and we expect that the corresponding results
 for ordinary GS superstring \cite{BK} will be helpful in this aspect.

 \section{An interplay between the nonabelian superstrings and the
 superstrings in a general $N=2$ supergravity background}
 \setcounter{equation}0

 It is interesting to compare our nonabelian superstring models
 (for the particular case $d=10, \; D=16$)
 with the models of superstrings moving in a curved 10-d supergravity
 background. We will see that our models provide special solutions
 for $N=2 \; 10-d$ supergravity.

 First of all, let us bring the action
 (\ref{u11}) into the general form of the action for the superstring
 evaluating in an arbitrary curved $N=2 \; 10-d$ superspace \cite{WIT,G,AB}
 \begin{eqnarray}
 A = \int d^{2} \xi \{ &-& \frac{1}{2} \eta_{\mu \nu} \sqrt{-g} g^{ab}
 (\partial_{a} z^{M}E^{\mu}_{M})
 (\partial_{b}z^{N}E^{\nu}_{N}) \Phi (z(\xi))  \nonumber  \\
 &+& \frac{1}{2} \varepsilon^{ab} \partial_{a}z^{M} \partial_{b}z^{N}
 B_{MN} \} = - \int d^{2} \xi {\cal L}(z^{M},\partial_{a}z^{N}) \label{w1}
 \end{eqnarray}
 Here $\Phi(z) $ is the dilaton superfield, which
 appears in the superfield formulation of the type II
 supergravities \cite{G,AB,GV};
 $\eta_{\mu \nu }$ is the flat 10-dimensional metric,
 $z^{M}=\{ x^{\mu}, \theta^{1\alpha}, \theta^{2\alpha} \}$ are the coordinates
 of the coset space $G_{1} \otimes G_{2}/G_{+}$
 (we can fix the gauge freedom by the
 condition $x^{1\mu}+x^{2\mu}=0$ and then put
 $x^{\mu}=x^{1\mu}-x^{2\mu}$) ; and $E^{\mu}_{M}=(\Phi)^{-1/2}(E^{1\mu}_{M}-
 E^{2\mu}_{M})$, where
 \begin{equation}
 \omega_{a}^{jA'}=\partial_{a} z^{jM'}E^{jA'}_{M'}=
 \partial_{a} z^{M}E^{jA'}_{M}    \label{w2a}
 \end{equation}
 (see eq. (\ref{u8})). Here and below the primed
 superindices (M', N', ...; A', B', ...) are the "$N=1$" ones used to
 numerate the parameters of the groups $G_{1}$ or $G_{2}$
 while the superindices without primes are the "$N=2$" ones, which
 correspond to the coordinates
 of the coset space $G_{1} \otimes G_{2}/G_{+} $.
 The extra superfield
 $B_{MN}$ possesses the symmetry properties $B_{MN}=
 -(-)^{MN}B_{NM}$ and in our case has the following explicit form
 \begin{eqnarray}
 B_{MN} &=& ((-)^{MN}E^{1\mu}_{M}E^{2\nu}_{N}-E^{1\mu}_{N}E^{2\nu}_{M})
 \eta_{\mu \nu}  \nonumber \\
 &+& 1/3(-)^{(C'+M)N} t_{A'B'C'}(E^{1C'}_{M}E^{1B'}_{N}B^{1A'}-
 E^{2C'}_{M}E^{2B'}_{N}B^{2A'})  \label{w2}
 \end{eqnarray}
 Nonlocal fields $B^{jA'}$ and supervielbeins $E^{jA'}_{N}$ have been defined
 in eqs. (\ref{u10}), (\ref{w2a}). Let us
 also introduce the two-form $B$ related
 to the tensor $B_{MN}$
 \begin{eqnarray}
 B &=& dz^{M} dz^{N} B_{MN}=  \nonumber  \\
 = 2E^{1\mu }E^{2\nu }\eta_{\mu \nu } &+& 1/3t_{A'B'C'}
 (E^{1C'}E^{1B'}B^{1A'}-E^{2C'}E^{2B'}B^{2A'})   \label{w3} \\
 dz^{M}dz^{N} &=& -(-)^{MN} dz^{N}dz^{M}  \label{w3b}
 \end{eqnarray}
 Here we deal with the one-forms $E^{jA'}=dz^{M}E^{jA'}_{M}$.
 Then we can define a closed
 three-form $H$ by
 \begin{equation}
 H=dz^{M}dz^{N}dz^{K}H_{MNK}=E^{C}E^{B}E^{A}H_{ABC}=d (B)  \label{w4}
 \end{equation}
 where
 \begin{equation}
 E^{A} \equiv ((\Phi )^{-1/2}(E^{1\mu}-E^{2\mu}),
 E^{1\alpha},E^{2\beta}),\; E^{A}E^{B} = (-)^{AB+1}E^{B}E^{A} \label{w4a}
 \end{equation}
 Using formulas (\ref{u6a}), (\ref{u9}), (\ref{u10}),
 (\ref{w2a}) and (\ref{w3}) we
 obtain for the three-form $H$ the following representation
 (it is useful to rewrite (\ref{u9}) as
 $dE^{jA'}=1/2E^{jC'}E^{jB'}t_{B'C'}^{A'})$
 \begin{eqnarray}
 H &=& (E^{1C'}E^{1B'}E^{2\nu }-E^{1\nu }E^{2C'}E^{2B'})
 t_{B'C'}^{\mu }\eta_{\mu \nu}+ \nonumber  \\
 &+& 1/3t_{A'B'C'}(E^{1C'}E^{1B'}E^{1A'}-E^{2C'}E^{2B'}E^{2A'})= \nonumber  \\
 &=& 1/3 \Phi^{3/2}t_{\mu \nu \lambda}E^{\lambda}E^{\nu}E^{\mu}+
 \Phi^{1/2} \Gamma_{\mu ,\alpha \beta}(E^{\mu}E^{2\alpha}E^{2\beta}+
 E^{\mu}E^{1\alpha}E^{1\beta}) \label{w5}
 \end{eqnarray}
 Here $\Gamma_{\mu ,\alpha \beta}=t_{\mu \alpha \beta}=-\eta_{\mu \nu}
 t^{\nu}_{\alpha \beta}$. For the components of $H$ we have
 \begin{eqnarray}
 H_{\mu \nu \lambda} &=& 1/3 \Phi^{3/2} t_{\mu \nu \lambda}  \; \; , \; \;
 H_{\mu (\alpha 1) (\beta 1)}=H_{\mu (\alpha 2) (\beta 2)}=
 \Phi^{1/2} \Gamma_{\mu , \alpha \beta}  \nonumber \\
 H_{\mu \nu (\alpha j)} &=& H_{(\alpha i)(\beta j)(\gamma k)}=
 H_{\mu (\alpha 1)(\beta 2)}=0   \label{w6}
 \end{eqnarray}
 The remaining components vanish. Let us now recall the
 definition of the torsion
 \begin{eqnarray}
 T^{C}_{AB} &=& (E^{M}_{A}\partial_{M} E^{N}_{B}-(-)^{AB}E^{M}_{B}\partial_{M}
 E^{N}_{A})E^{C}_{N}= \nonumber  \\
 &=& (-)^{NB}E^{N}_{A}E^{M}_{B}(\partial_{M}E^{C}_{N}-(-)^{MN}
 \partial_{N}E^{C}_{M}) \label{w7}
 \end{eqnarray}
 The factor $(-)^{NB}$ in (\ref{w7}) ensures the
 symmetry properties $T^{C}_{AB}=(-)^{1+AB}T^{C}_{BA}$. Substituting
 (\ref{u9}), (\ref{w4a}) into (\ref{w7}) we get
 \begin{eqnarray}
 T^{\mu}_{AB} &=& (-)^{N(B+C')}E^{N}_{A}E^{M}_{B} \Phi^{-1/2}
 (E^{1C'}_{M}E^{1B'}_{N}-E^{2C'}_{M}E^{2B'}_{N})t^{\mu}_{B'C'}+ \nonumber  \\
  &+& (-)^{NB}E^{N}_{A}E^{M}_{B}((-1/2)\Phi^{-1}\partial_{M}\Phi E^{\mu}_{N}+
  (-)^{MN}1/2 \Phi^{-1}\partial_{N}\Phi E^{\mu}_{M}) \label{w8a} \\
  T^{(\alpha i)}_{AB} &=& (-)^{N(B+C')}E^{N}_{A}E^{M}_{B}E^{iC'}_{M}
  E^{iB'}_{N}t^{\alpha}_{B'C'}  \label{w8b}
 \end{eqnarray}
 Introducing the supermatrices $e_{B}^{(iC')}=(E^{M}_{B}E^{iC'}_{M})
 \; (i=1,2)$
 we can rewrite (\ref{w8a}), (\ref{w8b}) in the form
 \begin{eqnarray}
 T^{\mu}_{AB} & = & (-)^{A(B+C')}t^{\mu}_{B'C'} \{ e^{(1C')}_{B}e^{(1B')}_{A}-
 e^{(2C')}_{B}e^{(2B')}_{A} \} \Phi^{-1/2} +   \nonumber \\
 & + & 1/2(\delta^{\mu}_{B} E^{M}_{A} \partial_{M} \ln \Phi -
 \delta^{\mu}_{A} E^{M}_{B} \partial_{M} \ln \Phi )  \label{w9}  \\
 T^{\alpha i}_{AB} & = & (-)^{A(B+C')}t^{\alpha}_{B'C'}
 e^{(iC')}_{B}e^{(iB')}_{A}.   \label{w9a}
 \end{eqnarray}
 We stress here once more that the multi-indices $(iC')$ and $(iB')$ run over
 $(\mu , \alpha i)$, for each $i$, while $A,B,...$ over
 $(\mu , (\alpha 1), (\beta 2) )$. This means that the supermatrices
 $e^{(iC')}_{B}$ are rectangular. Taking into account the
 definition of $E^{M}_{B}$ as inverse to
 $E^{A}_{M}=(\Phi^{-1/2}(E^{1\mu}_{M}-E^{2\mu}_{M}), E^{1\alpha}_{M},
 E^{2\alpha}_{M})$ we find
 \begin{eqnarray}
 e^{(1\mu)}_{A} &-& e^{(2\mu)}_{A}=\delta^{\mu}_{A} \Phi^{1/2}  \nonumber \\
 e^{(i\alpha)}_{(\beta j)} &=& \delta^{\alpha}_{\beta}
  \delta^{i}_{j} \; , \; e^{(i\alpha)}_{\mu}=0  \label{w10}
  \end{eqnarray}
 Therefore the supermatrices $e^{(iC')}_{B}$ are not independent
 and we can express the torsion components (\ref{w9}), (\ref{w9a}) in
 terms of a single supermatrix,
 for example in terms of $e^{(1C)}_{A} \equiv e^{C}_{A}$.
 Thus, for the components of torsion we obtain from eqs. (\ref{w9}),
 (\ref{w9a}) and (\ref{w10})
 \begin{eqnarray}
 T^{\mu}_{\nu \lambda} &=& t^{\mu}_{\nu ' \lambda '}(e^{\nu '}_{\nu}
 \delta^{\lambda '}_{\lambda} + e^{\lambda '}_{\lambda}\delta^{\nu '}_{\nu} )
 +\Phi^{1/2}t^{\mu}_{\nu \lambda} +  \nonumber \\
 &+& 1/2(\delta^{\mu}_{\lambda} E^{M}_{\nu} -\delta^{\mu}_{\nu}E^{M}_{\lambda})
 \partial_{M} \ln \Phi , \label{w11a} \\
 T^{\mu}_{\nu (\alpha 1)} &=& t^{\mu}_{\nu \lambda '}e^{\lambda '}_{(\alpha 1)}-
 1/2\delta^{\mu}_{\nu} E^{M}_{(\alpha 1)}\partial_{M} \ln \Phi , \label{w11} \\
 T^{\mu}_{\nu (\alpha2)} &=& t^{\mu}_{\nu \lambda '}e^{\lambda '}_{(\alpha 2)}-
 1/2 \delta^{\mu}_{\nu}E^{M}_{(\alpha 2)}\partial_{M} (\ln \Phi),
 \label{w11b} \\
 -T^{\mu}_{(\alpha 1)(\beta 1)} &=& T^{\mu}_{(\alpha 2)(\beta 2)}=
 \Gamma^{\mu}_{\alpha \beta} \; , \;
 T^{\mu}_{(\alpha 1)(\beta 2)}=0, \label{w11c} \\
 T^{(\alpha 1)}_{\mu \nu} &=& 0 \; , \; T^{(\alpha 1)}_{(\beta 2) \nu}=0 \; ,
 \; T^{(\alpha 1)}_{(\beta 2) (\gamma 2)}=0,  \label{w11d}  \\
 T^{(\alpha 1)}_{(\beta 1) \mu} &=& t^{\alpha}_{\beta \mu '}
 e^{\mu '}_{\mu } , \label{w11e}    \\
 T^{(\alpha 1)}_{(\beta 1) (\gamma 1)} &=&  t^{\alpha}_{\mu \gamma}
 e^{\mu }_{(\beta 1)}- t^{\alpha}_{\beta \mu }e^{\mu }_{(\gamma 1)} ,
 \label{w11f}  \\
 T^{(\alpha 1)}_{(\beta 1) (\gamma 2)} &=& -t^{\alpha}_{\beta \mu }
 e^{\mu }_{(\gamma 2)} . \label{w11g}
 \end{eqnarray}
 The remaining components
 have a similar form. Comparing (\ref{w11c})
 with the constraints on the torsion
 components $T^{\mu}_{(\alpha i)(\beta j)}$
 of the type $IIA$ and $IIB$ supergravities \cite{G,AB,GV} we conclude that
 our non-abelian superstring model belongs to
 the type $IIA$. We remind here that
 for the type $IIA$ we have the constraints like (\ref{w11c})
 while for the type $IIB$ \cite{G,AB,GV} the following ones
 \begin{equation}
 T^{\mu}_{(\beta i) (\alpha i)}=0 \; , \; T^{\mu}_{(\alpha 1)(\beta 2) } =
 T^{\mu}_{(\beta 2)(\alpha 1)}=\Gamma^{\mu}_{\alpha \beta} \; . \nonumber
 \end{equation}

  \section{Conclusion}

In this paper we have constructed the hamiltonian
formulation for the nonabelian N=2 superstring models proposed in
\cite{II}. This allowed us to deduce some elements of the corresponding
quantum theory. It seems that the complete quantum consideretion
will require the whole arsenal of the methods worked out for the
covariant quantization of Green-Schwarz superstring \cite{GS,KRNP,BK}.
In particular, the relevant ghosts and harmonic variables
are to be introduced (see \cite{KRNP,BK}).

In refs. \cite{WIT,G,AB,GV} ,
it has been shown, within the Lagrangian approach,
that the requirement of $ \kappa $-supersymmetry in the sigma
models of the type (\ref{w1}) leads to the constraints on the supergravity
background which are precisely the same as those imposed in the
standard superspace formulation of 10-d supergravity.
Using our approach one can easily derive an analogous result in the
framework of the Hamiltonian formalism.

It could be very interesting to apply
our methods to the case of nonabelian superstring
model based on the supergroup $ Osp(2 \mid 1) $, the structure constants
of which satisfy (\ref{u19}).
This model
can be interpreted as a Green-Schwarz superstring moving in
3-dimensional curved superspace which is coset space
$ Osp(2 \mid 1) \otimes Osp(2 \mid 1)^{*} / Sp(2) $.

 {\bf Acknowledgements}
We would like to thank Professor Abdus Salam, the International
Atomic Energy Agency and UNESCO for hospitality at the International
Centre for Theoretical Physics, Trieste where this work was completed.
We would also like to thank Professors E.Sezgin and J.A.Strathdee
for useful discussions.


 \end{document}